\documentclass[conference]{IEEEtran}
\IEEEoverridecommandlockouts
\usepackage{amsmath,amssymb,amsfonts}
\usepackage{algorithmic}
\usepackage{graphicx}
\usepackage{textcomp}
\usepackage{xcolor}

\usepackage{hyperref} 

\renewcommand{\figurename}{Fig.} 
\def\BibTeX{{\rm B\kern-.05em{\sc i\kern-.025em b}\kern-.08em
    T\kern-.1667em\lower.7ex\hbox{E}\kern-.125emX}}
\begin{document}

\title{Versatile CMOS Analog LIF Neuron for Memristor-Integrated Neuromorphic Circuits\\

\thanks{We acknowledge financial support from the EU: ERC-2017-COG project IONOS (\# GA 773228) and the CHIST-ERA UNICO project. FA thanks the MEIE for support through the chair in neuromorphic engineering.}
}
\author{
\IEEEauthorblockN{Nikhil Garg}
\IEEEauthorblockA{\textit{3IT-LN2, Université de Sherbrooke} \\
Sherbrooke, Canada \\}
\IEEEauthorblockA{\textit{IEMN-CNRS, Université de Lille} \\
Lille, France \\}
\and
\IEEEauthorblockN{Davide Florini}
\IEEEauthorblockN{Patrick Dufour}
\IEEEauthorblockA{\textit{3IT-LN2, Université de Sherbrooke} \\
Sherbrooke, Canada \\}
\and
\IEEEauthorblockN{Eloir Muhr}
\IEEEauthorblockN{Mathieu C. Faye}
\IEEEauthorblockN{Marc Bocquet}
\IEEEauthorblockA{\textit{IM2NP-CNRS,
Aix-Marseille Université} \\
Marseille, France \\}
\and
\IEEEauthorblockN{Damien Querlioz}
\IEEEauthorblockA{\textit{C2N-CNRS,
Université Paris-Saclay} \\
Paris, France \\}
\and
\IEEEauthorblockN{Yann Beilliard}
\IEEEauthorblockN{Dominique Drouin}
\IEEEauthorblockA{\textit{3IT-LN2, Université de Sherbrooke} \\
Sherbrooke, Canada \\}
\and 
\IEEEauthorblockN{Fabien Alibart}
\IEEEauthorblockA{\textit{3IT-LN2, Université de Sherbrooke} \\
Sherbrooke, Canada \\}
\IEEEauthorblockA{\textit{IEMN-CNRS, Université de Lille} \\
Lille, France \\}
\and
\IEEEauthorblockN{Jean-Michel Portal}
\IEEEauthorblockA{\centerline{\textit{IM2NP-CNRS,
Aix-Marseille Université}} \\
Marseille, France \\}

}

\maketitle

\begin{abstract}
Heterogeneous systems with analog CMOS circuits integrated with nanoscale memristive devices enable efficient deployment of neural networks on neuromorphic hardware. CMOS Neuron with low footprint can emulate slow temporal dynamics by operating with extremely low current levels. Nevertheless, the current read from the memristive synapses can be higher by several orders of magnitude, and performing impedance matching between neurons and synapses is mandatory. In this paper, we implement an analog leaky integrate and fire (LIF) neuron with a voltage regulator and current attenuator for interfacing CMOS neurons with memristive synapses. In addition, the neuron design proposes a dual leakage that could enable the implementation of local learning rules such as voltage-dependent synaptic plasticity. We also propose a connection scheme to implement adaptive LIF neurons based on two-neuron interaction. The proposed circuits can be used to interface with a variety of synaptic devices and process signals of diverse temporal dynamics.
\end{abstract}


\section{Introduction}

Analog circuits and devices can efficiently emulate neural dynamics in real-time by utilizing physical mechanisms such as Kirchoff's law through memristive synapses for signal transmission and capacitive charging on CMOS devices for temporal integration of current \cite{joubert2012hardware}. Such strategies are inherited from the seminal work of C. Mead~\cite{Mead1990} to implement silicon neurons with rich temporal dynamics at low power. More recently, nanoscale non-volatile memristive devices have been considered for synaptic function implementation and offer several interesting features, such as non-volatile synaptic weight storage, analog programming, and low-power operation. Such devices are also CMOS compatible and could allow 3D integration \cite{shulaker2017three} for high-density synaptic arrays to implement in-memory computing architectures \cite{sebastian2020memory}. The development of spiking neural networks based on these technologies still needs to address several challenges, such as synapses/neuron impedance matching and operation stability validation. 

\begin{figure*}[ht]
    \centering
    \includegraphics[width=1\textwidth]{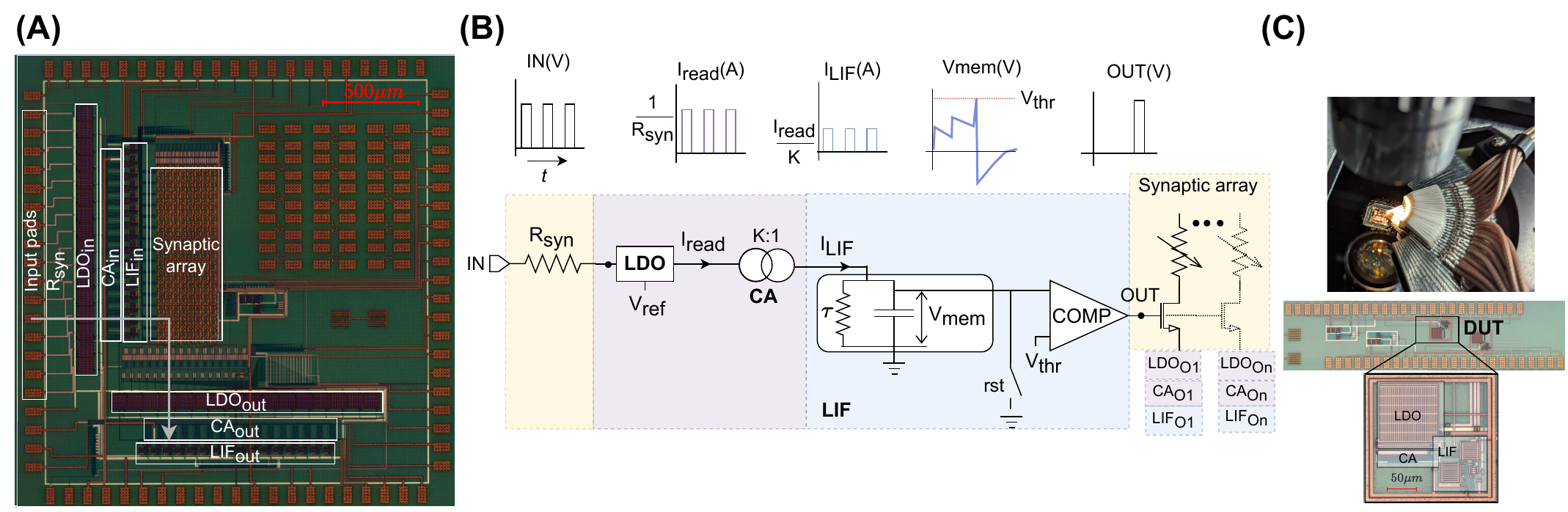}
    \caption[Proposed architectures and approach]{(A) Optical micrograph of the realized "UNICO" ASIC with an arrow denoting the signal chain from IO pad to neurons of the input layer ($LIF_{in}$) through a resistive synaptic column and then to output layer ($LIF_{out}$) through a memristive synaptic array. (B) The signal chain for spike transmission consists of a synaptic resistance, a low dropout regulator (LDO) for regulating the voltage at other synaptic terminals, and a current attenuator for down-scaling the current by a factor of 'K'. The leaky integrate and fire (LIF) neuron integrates the read current. The neuron membrane potential tracks the neuron state and is reset when crossing a threshold voltage level. Consequently, a spike is transmitted to the next layer of neurons through the synaptic array. (C) Test structures with connection pads probed for electrical characterization of implemented circuit blocks. 
}  
    \label{fig1}
\end{figure*}

This study presents an architecture and circuits that enable stable reading of the memristive synapse into an analog spiking neuron circuit. These circuit blocks, including a low-dropout regulator (LDO) and current attenuator, were implemented alongside a LIF neuron in a 130nm CMOS integrated circuit (IC). The measurement results demonstrate the sensitivity of the neuron's activity for a range of synaptic resistance, excitation voltage levels, and pulse widths representative of the memristive synapse operation. The modulation of threshold, leak rate, and refractory period showcase the configurability of neuron dynamics that could be adapted to match different synaptic array sizes, device conductance range, and the time scale of the deployed application. Finally, an architecture to configure generic memristive LIF neurons to an adaptive variant is proposed.

Neurons with slow leak rates and long-time-scale dynamics can be critical for emulating bio-realistic dynamics in real time. Specialized low-pass filtering circuits (\cite{Chicca2014NeuromorphicEC}, \cite{Indiveri2003ISCAS}) are often used to implement such slow dynamics with small capacitance. We implemented such an integrator circuit with modifications to realize controllable bi-directional leakage for local learning with voltage-dependent synaptic plasticity (VDSP) \cite{garg2022voltage, goupy2023unsupervised}. In contrast to previous studies \cite{Wu2015neuron, Lecerf2014ISCAS}, the interfacing between the integrator operating at low-current levels and the memristive device was enabled by a current attenuator implemented in the signal chain. VDSP-based synaptic plasticity can further exploit such dynamics to detect spike patterns in longer time windows and enable efficient local learning. 

\begin{figure*}[hbt]
    \centering
    \includegraphics[width=1\textwidth]{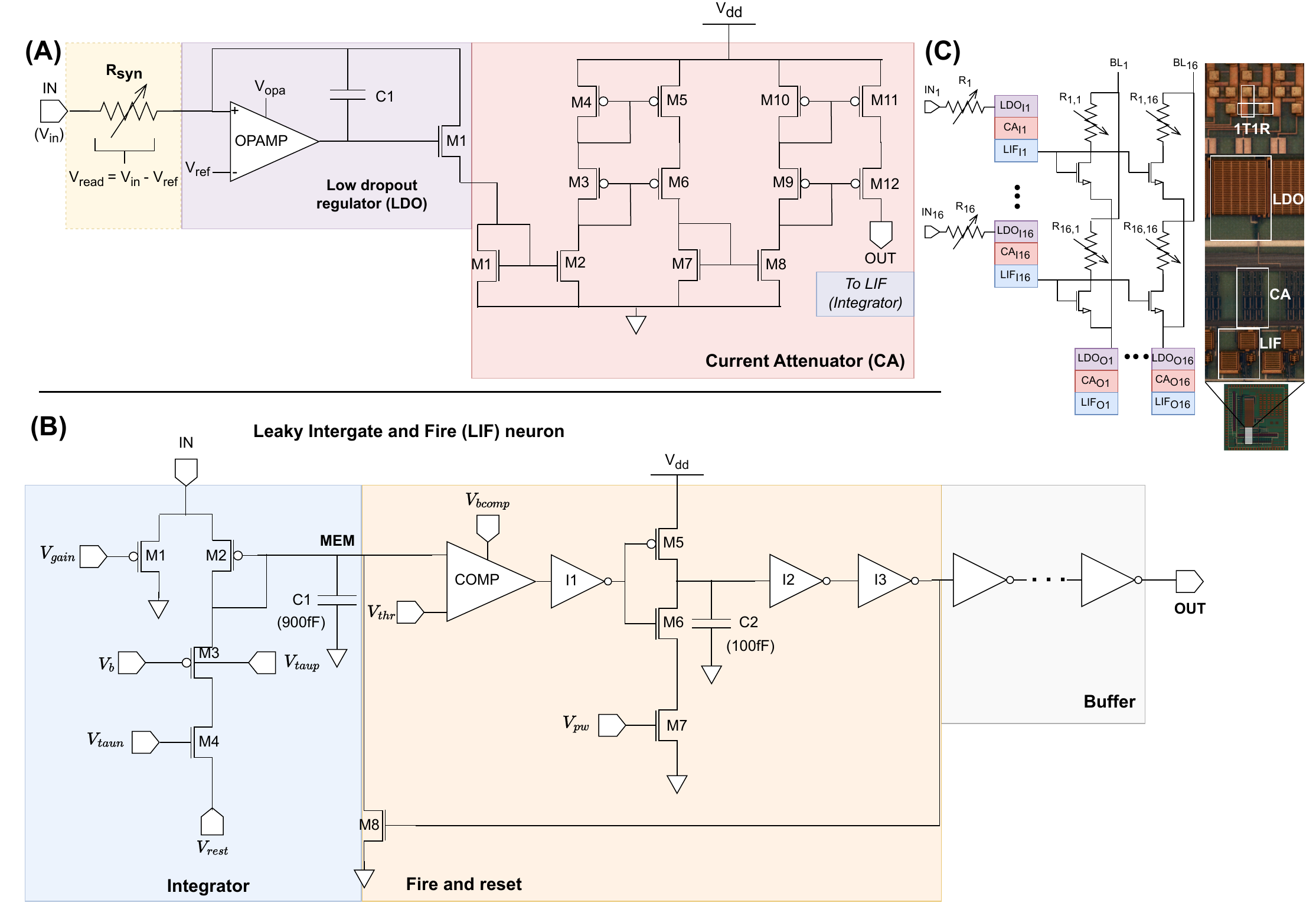}
    \caption[Circuits]{Key circuit blocks. (A) This schematic shows the synaptic resistance, Low Dropout Regulator (LDO), and current attenuator arranged to read the synaptic current to the neuron. 
    (B) The LIF neuron is depicted as comprising a leaky integrator, fire and reset block, and buffer. (C) The schematic displays the synaptic array architecture on the left, and on the right, a micrograph of the realized chip is zoomed in around the signal chain from the synaptic array to the neuron.
    }
    \label{Circuits}
\end{figure*}

The manuscript is organized as follows. The Methods section details the realized ASIC architecture with key circuit blocks to illustrate the signal chain. The results section presents the electrical characterization of the neuron subject to various input synaptic currents. The membrane threshold level, and refractory period impact on spiking activity is further evaluated. Next, the characterization results in response to spike-based (pulsed) stimuli are presented with an analysis of leak rate modulation. In the end, a connection topology for the dynamic configuration of generic LIF neurons to adaptive LIF neurons is presented, and hardware overheads for different circuit functionalities are discussed.

\section{Materials and Methods}

\subsection{Implemented Design}

The application-specific integrated circuit (ASIC) was realized for hardware implementation of SNN on a monolithic chip through CMOS analog neurons and back end of the line (BEOL) integrated memristive~\cite{el2022fully} synaptic devices (\autoref{fig1}(A). The memristive devices are integrated in this design in a 1T1R configuration. In such a configuration, the output terminal of the pre-neuron sends voltage spikes (pulses) to the gate of the 1T1R downstream synaptic array that will enable input current to the post-neurons. The signal chain of the synapse and neuron is illustrated in \autoref{fig1}(B). For neuron characterization, the presynaptic input is a voltage applied to the resistor's first terminal (IN), emulating the memristive synapse. LDO clamps the second terminal at $V_{ref}$, and the $I_{syn}$ current weighted by the resistance $R_{syn}$ is scaled by a factor of K by the current attenuator. The LIF neuron membrane voltage performs temporal integration of the current and resets to $V_{reset}$ when $V_{mem}$ is higher than the Vth threshold voltage. Test structures with connection pads implemented on the ASIC were probed for testing (\autoref{fig1}C). The analog bias voltages and input signals were generated through an FPGA-controlled PCB, and signals from ASIC were measured through an oscilloscope.

The full SNN integrates two layers of neurons in which the output spike is transmitted to the gates of all the downstream synapses of the memristive array. The output of the synaptic array is connected to the second layer of LIF neurons, which present again the successive blocks of LDO and current attenuator.
\begin{table}
\caption{Bias voltages used in circuits}
\begin{center}
\begin{tabular}{|l|l|l|}
\hline
Name & Typical (V) & Purpose \\ \hline
$V_{dd}$               &     3.3              &   Power supply          \\ \hline
$V_{ref}$        &             2.4      &       LDO reference      \\ \hline
$V_{opa}$        &             2.4      &       OPAMP bias (LDO)      \\ \hline
$V_{gain}$        &                   2.1&      Gain modulation       \\ \hline
$V_{taun}$               &           1.2        &  Leak rate (Down)\\ \hline
$V_{taup}$        &                   1.2&         Leak rate (UP)   \\ \hline
$V_{rest}$               &                   0.6&    Resting potential         \\ \hline
$V_{thr}$        &                 1.2  &    Neuron threshold level        \\ \hline
$V_{bcomp}$               &        2.4           &   Bias for COMP (LIF) \\ \hline
$V_{pw}$               &             1      &     Pulse width modulation      \\ \hline 
\end{tabular}
\label{bias}
\end{center}
\end{table}

\begin{figure*}[ht]
    \centering
    \includegraphics[width=1\linewidth]{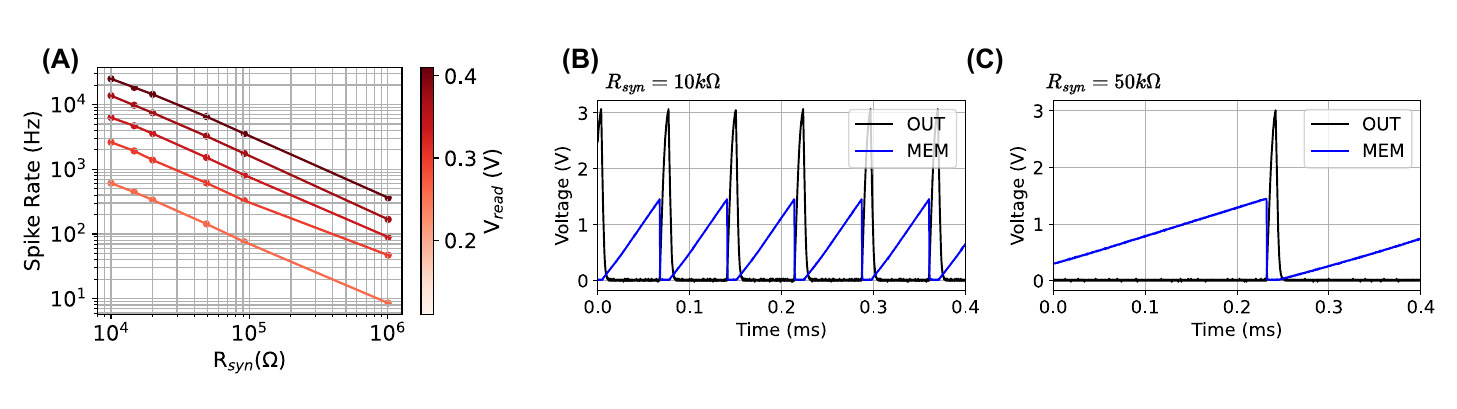}
    \caption{Characterization of neuron's sensitivity to synaptic resistance.
    (A) Measured output spike rate of neuron with respect to $R_{syn}$ for different $V_{read}$ levels between 100mV and 400mV. The respective $V_{read}$ was applied for 1s for each experiment, and the neuron's response was recorded. Comparison of neuron's membrane voltage and output response is shown for $R_{syn}$ of 10k$\Omega$ (B) and 50k$\Omega$ (C) with read voltage ($V_{read}$) of 250mV. 
    }
    \label{RES}
\end{figure*}
\subsection{Circuits}

The circuit schematics of LDO and current mirror are shown in \autoref{Circuits}(A). External Input voltage ($V_{in}$) is applied to the first terminal of the synaptic resistance ($R_{syn}$). The other terminal of the synapse is connected to the (+) input terminal of the operational amplifier (OPAMP) in LDO. The other terminal (-) of OPAMP is connected to $V_{ref}$. Whenever a voltage greater than $V_{ref}$ is applied on IN, the output of OPAMP rises, and transistor M1 is turned on. The current read from $R_{syn}$ is transmitted via the feedback branch, and the voltage at the positive input of OPAMP is pulled to $V_{ref}$. The output current of LDO is the accumulation of weighted current from all upstream synapses, which is fed to the downstream current attenuator.  

The current attenuator is implemented with a cascoded current mirror. This topology is beneficial for a wide swing \cite{bruun1995dynamic} across the resistance range of upstream memristive synapse. The attenuation factor was tuned by adjusting the W/L ratio of each transistor pair (e.g., M1 and M2, M3 and M4, etc.). Additionally, the sizing was performed to ensure a constant attenuation factor of 500 throughout the read current range. Such topology of LDO and current mirror is also referred to as active current mirror in previous studies \cite{miguez2017active}. 

The LIF neuron (\autoref{Circuits}B) performs temporal integration of signals from the current attenuator and consists of a low pass filtering based on difference pair integrator (DPI) circuit \cite{Indiveri2003ISCAS} for implementing the integration of synaptic current on the capacitor ($C1$). The resting state potential bias voltage controls the level to which the membrane potential leaks ($V_{rest}$) and was set higher than the reset potential (0V). A transistor (M3) was added to the DPI circuit to realize bi-directional leakage of membrane potential, and the leak rate is controlled by $V_{taup}$ and $V_{taun}$ bias voltages.

The fire and reset block is composed of a comparator for membrane voltage threshold crossing detection with an externally supplied threshold level ($V_{thr}$). This block is biased by $V_{biascomp}$. The neuron's pulse width and refractory period are configured by modulating the discharging rate of capacitance $C2$ through bias voltage $V_{pw}$. A reset transistor (M8) discharges the membrane capacitor to the ground on a spike event, and the generated spike is transmitted to the downstream synapse through a buffer. As the reset transistor is activated throughout the spike generation period, $V_{pw}$ also controls the neuron's refractory period. 

The architecture of the memristive synaptic array is shown in \autoref{Circuits}(C)(left). The neurons in the input layer ($LIF_{I1}$ to $LIF_{I16}$) transmit a generated spike to the next layer by applying the spike output to the transistor’s gate of the 1T1R synaptic cell. The spike amplitude is set to $V_{dd}$ to minimize the voltage drop access transistor during spike transmission. All the synaptic cells in a column are connected to the corresponding output neuron through LDO and the current attenuator. The circuits were implemented in 130nm CMOS technology, and a picture from a microscope is shown to visualize the layout of the signal chain in \autoref{Circuits}(C)(right). The typical values and short descriptions of bias voltages are summarized in \autoref{bias}.

 \begin{figure*}[ht]
    \centering
    \includegraphics[width=1\linewidth]{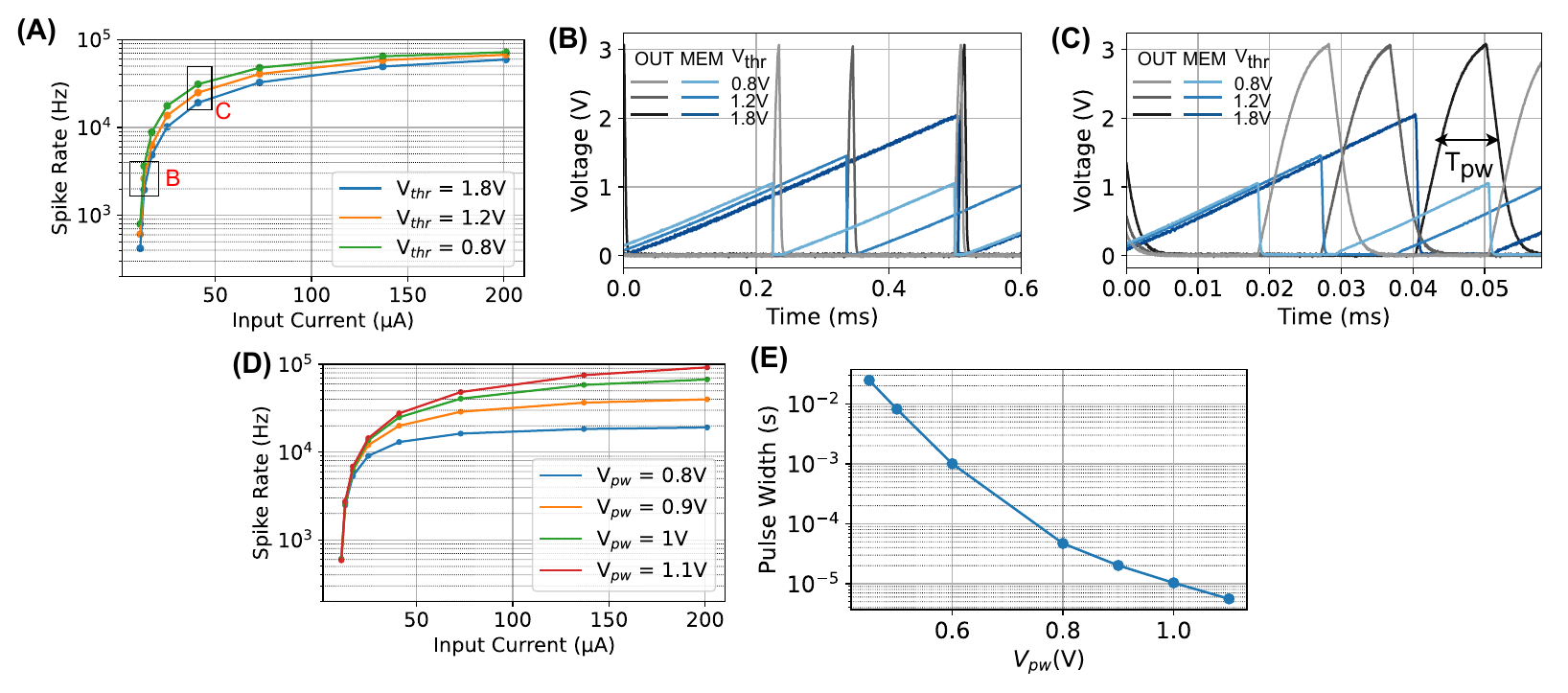}
    \caption{Characterization of neuron's transfer function with DC excitation. (A) Measured spike rate of neuron with respect to input current ($I_{syn}$) and spiking threshold bias level ($V_{thr}$). Comparison of the measured temporal response of membrane voltage and output for different threshold levels obtained with $I_{syn}$ of 10$\mu$A (B) and 40$\mu$A (C). (D) Neuron's spike rate with respect to input current for different $V_{pw}$ bias levels. (E) Pulse width ($T_{pw}$) of the generated output spike plotted with respect to $V_{pw}$ bias voltage.}
    \label{DC}
\end{figure*}

\section{Results and Discussion}

\subsection{Synaptic resistance reading}

For characterizing the synaptic resistance reading, an external resistor ($R_{ext}$) was connected in series with the on-chip resistor of 10k$\Omega$ ($R_{IC}$) to emulate a synaptic resistance, resulting in read current $I_{syn} = (V_{in}-V_{ref})/ (R_{ext}+R_{IC})$, where $V_{ref}$, the reference voltage supplied to LDO, was set to 1V, and $R_{ext} + R_{IC}$ represents the net synaptic resistance ($R_{syn}$).
The relationship between the measured output spike rate and input resistance ($R_{syn}$) is shown in \autoref{RES}(A). The input resistance was varied from 10k$\Omega$ to 1M$\Omega$, and experiments were repeated for five different read voltages ranging between 100mV and 400mV. Such resistance range matches the one observed in our memristive devices. The applied voltage is also compatible with the read voltage range from the memristive synapses (i.e., read without weight disturbance).

To illustrate the temporal response of the neuron, the measured membrane voltage and output of the neuron are compared in \autoref{RES}(B-C) for synaptic resistance of 10k$\Omega$ and 50k$\Omega$ and read voltage ($V_{read}$) of 250mV. 

Sensitivity to the large range of synaptic resistance makes the neuron suitable for various memristive technologies. In the above experiments, the output spike rate was observed to vary between 8Hz and 25kHz. The ability to read and differentiate between low resistances (with high firing rates) can enable reading from several presynaptic resistances in parallel.

\subsection{Neuron transfer characteristics}
\begin{figure}[h]
    \centering
    \includegraphics[width=0.3\textwidth]{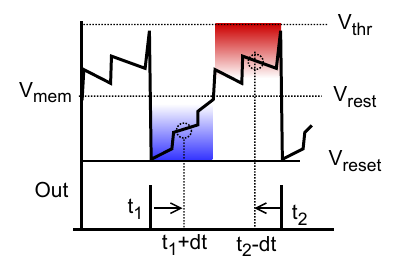}
    \caption{Illustrative plot to show the behavior of a neuron with bi-directional leakage. The downward and upward leakage occurs when membrane voltage is greater or less than resting state potential ($V_{rest}$), respectively. A leak in the upward direction occurs when the neuron is reset after the spike event ($t_1$) and enables estimation of time elapsed (in blue). Similarly, the occurrence of a spike event in the near future ($t_2$) can be predicted through a high value of membrane voltage (in red).}
    \label{VDSP}
\end{figure}
\begin{figure*}[ht]
    \centering
    \includegraphics[width=1\textwidth]{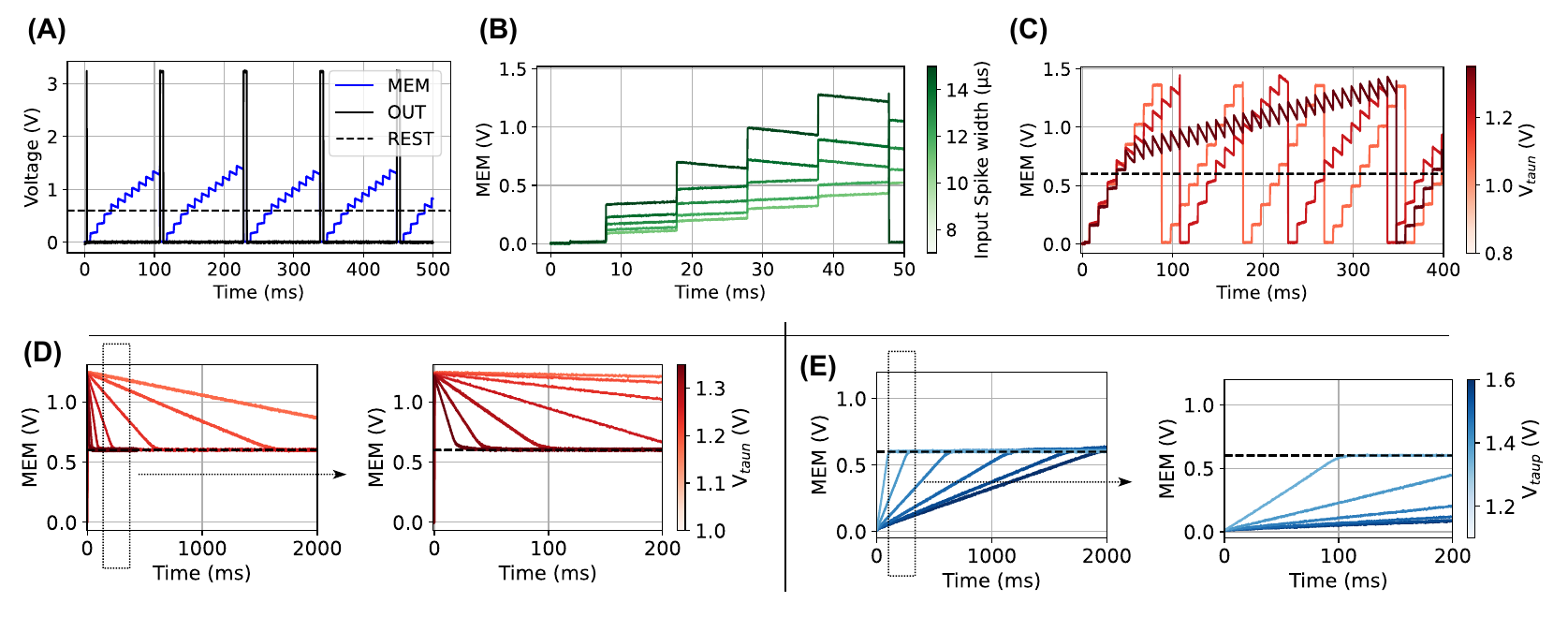}
    \caption{Characterization of temporal dynamics. (A) Input pulses of width 10$\mu$s were applied at 100Hz, with the neuron's membrane potential and output shown across time. (B) The input pulse width was varied between experiments (7$\mu$s and 15$\mu$s), comparing the charging events of membrane voltage. (C) The experiment varied $V_{taun}$ across experiments to compare the neuron's membrane voltage evolution across time. In (A-C), the $I_{syn}$ for each pulse was of magnitude 40$\mu$A. (D) A single input pulse was applied to charge the membrane voltage close to the neuron's threshold, and the downward leak rate is controlled by $V_{taun}$ bias level and varied across experiments. (E) The magnitude of the applied spike was increased to induce neuron firing. The leakage in the upward direction (from $V_{reset}$ to $V_{rest}$) is compared by varying $V_{taup}$  bias level between experiments.}
    \label{Spike}
\end{figure*}

To characterize the excitation of neurons for a range of synaptic currents and bias voltages, the excitation current was varied from 10$\mu$A to 200$\mu$A. The read voltage was applied through an on-chip resistor of 10k$\Omega$ for a long duration (100ms). The rate of output spikes generated by the neuron is plotted with respect to the injected current, and the neuron's threshold voltage level ($V_{thr}$) is shown in \autoref{DC}(A). For the threshold level of 1.8V, the spike rate was 419Hz and 59kHz for input current of 10$\mu$A and 200$\mu$A, respectively. Similarly, for a threshold level of 0.8V, the spike rate varied between 800Hz and 68kHz. The measured membrane voltage and output spike are compared for different threshold voltages and input currents of 10$\mu$A (B) and 40$\mu$A (C). 

The threshold impacts the range of firing rate (activity) that could be observed with respect to input current. A lower threshold could increase the sensitivity to low current (high synaptic resistance), but the spike rate saturate early. Conversely, a high threshold led to lower firing rates for small excitation currents, and a variation in spike rate could be observed even for high input currents.
The maximum spike rate of the neuron is limited by the output pulse width or refractory period controlled by $V_{pw}$ bias level. The neuron's response(rate) to different ($V_{pw}$) is shown in \autoref{DC}(D). The maximum firing rate was 20kHz and 92kHz for ($V_{pw}$) of 0.8V and 1.1V, respectively. The $V_{pw}$ bias voltage modulates the output spike's pulse width and the neuron's refractory period by controlling the discharging rate of the $C2$ capacitor in \autoref{Circuits}(C). The pulse width was measured as the time difference between the crossing of OUT and $V_{dd}/2$ and is plotted for $V_{pw}$ in \autoref{DC}(E). The resultant pulse width of the output spike varied between  20ms and 10$\mu$s for $V_{pw}$ of 0.45V and 1V, respectively.

\subsection{Temporal dynamics}

The neuron was characterized by applying short pulses at frequent intervals to access charging with respect to input pulse width and dynamics in the absence of excitation. Spikes of magnitude 40$\mu$A and width 10$\mu$s were fed to the neuron by applying voltage pulses ($V_{read}$=400mV) across the 10k$\Omega$ on-chip resistor at the rate of 100Hz. The measured response of the neuron membrane voltage and output spikes is plotted in \autoref{Spike}(A) and results in a neuron spike rate of 8Hz.
The above experiment was repeated for different pulse widths of input spikes between 5$\mu$s to 15$\mu$s. The measured response of the neuron's membrane voltage across time is plotted in \autoref{Spike}(B) to compare the magnitude of the increment in membrane voltage for every charging event. A pulse of 7$\mu$s led to a small increment in membrane voltage at every spike event, resulting in a firing rate of 2Hz. Whereas a 15$\mu$s input pulse could fully charge the neuron with 5 input spikes, resulting in an output spike rate of 18Hz.
Since the pulse width influences the number of spikes, the postsynaptic neuron integrates before firing the pulse width of the presynaptic neuron spikes, which can be adapted for the crossbar size. For example, suppose a neuron in the postsynaptic layer is expected to integrate signals from four times input neurons in parallel. In that case, the pulse width of input neurons should be accordingly reduced by increasing the $V_{pw}$ bias level to maintain similar temporal dynamics in the output layer. A shorter pulse also leads to a granular increment in membrane voltage, allowing down-sizing of the membrane capacitance. 
In the absence of input current, the membrane voltage leaks to a resting state potential ($V_{rest}$). Since the reset level was set to 0V, the neuron exhibits leakage in upward direction after the refractory phase to $V_{rest}$ (set to 600mV). This bi-directional mechanism was implemented to differentiate between idle neurons and those in the refractory period, as shown in \autoref{VDSP}, and is beneficial for implementing local synaptic learning. 

An important aspect of memristive synapses / CMOS neuron co-integration for SNN is to allow local learning based on the activity of the adjacent neurons  ~\cite{querlioz2013immunity}. In local learning rules such as Spike Timing Dependent Plasticity (STDP), the spike time difference between the pre and post-neurons is converted into a programming voltage by overlapping of slow decaying voltage pulses~\cite{Saïghi2015Frontiers}; however, this approach can consume a significant fraction of area-energy budget \cite{ambrogio2016unsupervised} and implementing pulses with complex shapes can become challenging. More recently, Voltage-dependent synaptic plasticity (VDSP) \cite{garg2022voltage, goupy2023unsupervised} was proposed to implement learning efficiently in hardware without requiring pulse shaping circuits. In this approach, the recent pre neuron’s activity can be estimated through the neuron membrane potential (i.e., low membrane potential being associated with a recent firing event and high membrane potential associated with an imminent spiking event). Mapping of this concept to memristive devices programming requires the implementation of more complex membrane dynamics. In this study, we present a neuron circuit with bi-directional leakage to enable learning of memristive weights through VDSP.

For characterizing the tunability of leak rate in the downward direction, spikes of magnitude 40$\mu$A and width 10$\mu$s were fed to the neuron at 100Hz, and $V_{taun}$ was varied between experiments. The measured membrane voltage response is compared in \autoref{Spike}(C). For $V_{taun}$ of 1.2V (typical), the output spike rate was 8Hz, which could be increased to 12Hz by lowering the leak rate. Conversely, increasing the leak rate could decrease the output spike rate to 2Hz. 

For evaluation of $V_{taun}$, the neuron was excited by a single input pulse of higher magnitude to charge the membrane voltage to just below the spiking threshold, and the response was recorded for 2s. The membrane voltage is compared for values of $V_{taun}$ between 1V and 1.35V in \autoref{Spike}(D) for time intervals of 2s and 200ms. The neuron could fully leak from the threshold level to resting state potential in 8 ms for a high leak rate. Lowering the leak rate could modulate this duration to more than 2s.

The leak rate impacts the neuron's memory window and can be adjusted to match the dynamics of the input signals. For instance, in scenarios where the neuron receives sparse spikes at a low frequency, reducing the leak rate can help preserve the neuron's memory over a large temporal window. Conversely, increasing the leak rate when processing high-frequency input signals can regulate the neuron's output firing rate.

Next, we characterized the upward modulation of the leak rate. The neuron was stimulated with a single high-magnitude spike to trigger an output spike event, and its membrane potential subsequently leaked from $V_{reset}$ to $V_{rest}$, as illustrated in \autoref{Spike}(E). The biasing voltage ($V_{taup}$) was adjusted between 1.1V and 1.6V. We monitored the membrane voltage over 2-second periods to assess the upward leak rate. The ability to tune the upward leak rate and the resting state voltage is beneficial for adjusting the learning dynamics by modifying the probabilities of potentiation and depression.
 \begin{figure*}[ht]
    \centering
    \includegraphics[width=1\linewidth]{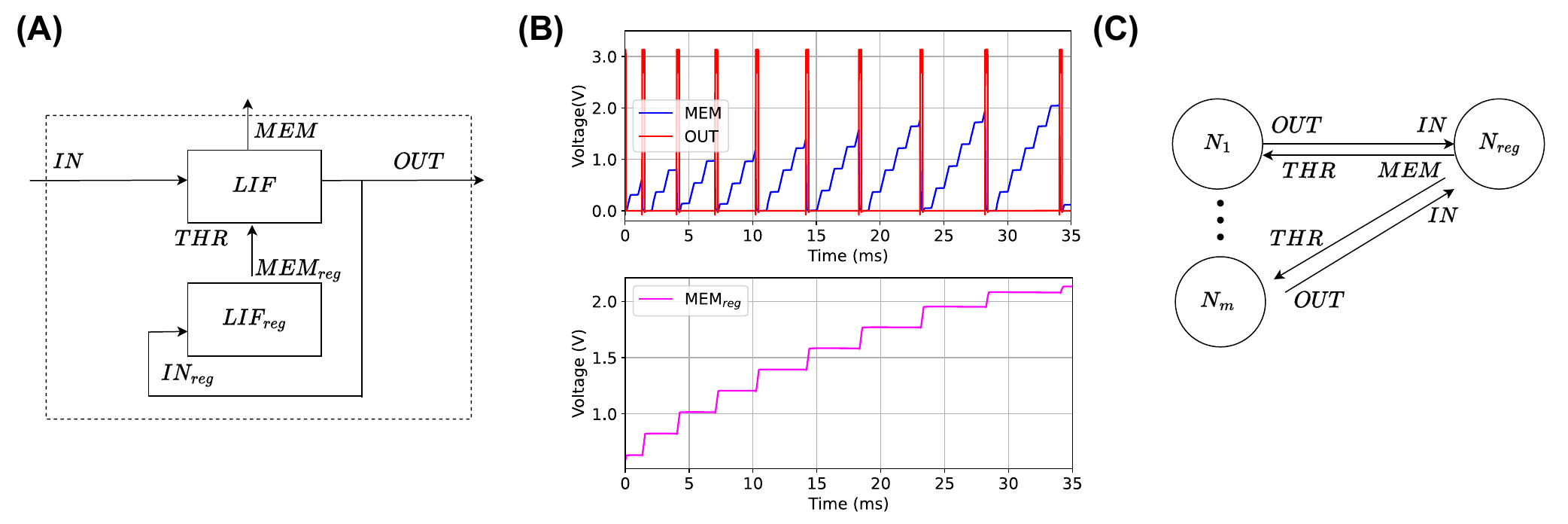}
    \caption[Caption]{
        (A) Connection scheme to realize an adaptive LIF neuron with a pair of LIF neurons. The regulator neuron integrates the output signals of the primary neuron, and its membrane voltage ($MEM_{reg}$) is used as the threshold of the primary neuron. 
        (B) The primary neuron was stimulated at a 1kHz rate and connected to a regulator neuron. The membrane potentials of both neurons and the output of the primary neuron are plotted using results from a SPICE circuit simulation.
        (C) Single regulator neuron ($N_{reg}$) shared by a sub-population of generic LIF neurons ($N_{1}, ..., N_{m}$). 
    }
    \label{Adap}
\end{figure*}
\subsection{Configurability to adaptive neuron}

The spiking neurons' information processing and memory capacity can be further enhanced through threshold adaptation~\cite {laughlin1981simple, naud2008firing, benda2003universal}. The designed neuron block can be used as a regulator neuron to monitor the spiking activity of a primary neuron. As depicted in \autoref{Adap}(A), the output of the primary neuron (OUT) is connected to the input terminal of the regulator neuron ($IN_{reg}$). This connection tracks the neuron activity through the membrane potential of the regulator neuron ($MEM_{reg}$), which in turn sets the threshold ($THR$) for the primary neuron.

The designed circuit blocks, connected as described in \autoref{Adap}, were simulated in SPICE. The primary neuron, excited by 1 kHz pulses, and the corresponding changes in membrane potential for both the primary and regulator neurons are depicted in \autoref{Adap}(B). During this simulation, the threshold voltage of the primary neuron ($MEM_{reg}$) increased from 600mV to 2V, resulting in the inter-spike interval of the output spikes increasing from 2ms to 5ms. A chain consisting of a 10k$\Omega$ synaptic resistance, followed by an LDO and a current attenuator, was used to transmit output voltage pulses from the primary to the regulator neuron. Alternatively, this synaptic resistance can be replaced with a memristor, which can be programmed to introduce an additional level of plasticity at the neuron level.

Additionally, a sub population of neurons ($N_1$, $N_2$, …, $N_m$) could share a common regulator neuron ($N_{reg}$), as illustrated in \autoref{Adap}(C).  Having a pool of LIF with the same regulator neuron can improve the system's scalability by sharing resources. This architecture can configure a fully analog LIF neuron chip to an arbitrary population of LIF and ALIF neurons in runtime. Such a heterogeneous population has been shown to enhance the learning capabilities of SNNs ~\cite{gutierrez2019population, salaj2021spike, patterson2013distinct}. Furthermore, monitoring neuron activity over time through traces can also enhance local learning \cite{mitra2008real}, providing a dynamic feedback mechanism to improve performance and adaptability.

\subsection{Overhead of different functionalities}

The power consumption of the implemented LDO and current attenuator was estimated through SPICE simulations of implemented circuits. The static power consumption, measured when no signal was present at the input, was 10µW for the LDO and 10pW for the current attenuator. During the operation of reading from the resistive synapse, the dynamic power consumption was estimated to be 18µW for the LDO and 20 µW for the current attenuator. The static power dissipation of LDO can be accounted for leakage due to the high biasing current used to obtain fast response time and high output current levels (up to 200µA). This quick activation is necessary when using short-reading pulses (neuron output spikes). As discussed in the earlier section, short pulses lead to gradual charging of (post-synaptic) neurons, thus helpful for scaling up crossbar size or the number of pre-synaptic neurons. Moreover, power dissipation through the memristive device during reading can be lower with short-read pulses. As these circuits are shared by synapses in a row or column, energy utilization can be balanced. 

The LIF neuron's static energy and energy per spike were estimated to be 5µW and 200 pJ/spike. The high static power can be attributed to leakage due to biasing the comparator. With a lower bias current, the static power can be reduced to 17nW, but the energy per spike increases to 7nJ. This trade-off occurs because a slow comparator can lead to high leakage during the firing phase. Feedback mechanisms \cite{nair2019ultra} can potentially reduce the energy per spike of the neuron with a low comparator bias current. 
Overheads associated with reading memristive states are also present in hardware Artificial Neural Network (ANN) implementations, where trans-impedance amplifiers and analog-to-digital converters (ADCs) sense the current from the synaptic column for subsequent computational processing and account for a major fraction of energy consumption. However, analog neurons are more suitable for interacting with analog memristive synapses. SNNs benefit from events' sparse activity, and gating the power supply \cite{chundi2021always} of synapse reading blocks around spike events can reduce static power dissipation through leakage.

\section{Conclusion}

We propose a versatile CMOS circuit to integrate memristive synapses into the signal chain of analog neuromorphic circuits. This integration pathway includes a Low-Dropout Regulator (LDO), a current attenuator, and an analog Leaky Integrate-and-Fire (LIF) neuron. The circuit blocks were implemented on a 130nm CMOS ASIC. The chip features multiple instances of these blocks alongside a synaptic array composed of 1T1R cells, facilitating the integration of memristive synaptic devices.
The circuit building blocks of the signal chain in the implemented ASIC were characterized. We demonstrated the neuron's sensitivity to the synaptic state by testing resistances ranging from 10k$\Omega$ to 1M$\Omega$. As a result, the neuron's firing rate was observed to vary between 8 Hz and 25 kHz. Additionally, the neuron's activity was characterized across a range of applied read voltages. By increasing the spiking threshold and pulse width of the neuron, the neuron can be fine-tuned to lower and higher firing rates, respectively.
Excitation with a sparse train of pulses was used to measure the neuron's temporal response. The neuron's charging was controllable by changing the input pulse width. Consequently, the presynaptic neuron spike's width can be adjusted per the postsynaptic neuron's fan-in characteristics. 
In the absence of an input signal, the neuron retains its memory through leakage, with an adjustable leak time constant. The bi-directional leakage enables the estimation of neurons' recent activity and should benefit online learning through local synaptic learning. 

We experimentally validated the modulation of the leak rate across short and long-term intervals, which can be helpful in processing signals with various temporal dynamics. The topology of pair neurons for configuration with adaptive neurons could further enhance the capabilities of the implemented neuron. Future work will detail the architecture and characterization of the integrated CMOS-RRAM ASIC with in-situ learning.

\section*{Author Contributions}

N.G. designed the circuits with contributions from E.M. and M.C.F., under the direction of J.M.P.; D.F. performed post-processing of the chip to expose the pads for probing. N.G. performed the on-chip experimental measurements, with contributions from P.D.; N.G. wrote the initial version of the manuscript. J.M.P., F.A., D.D., Y.B., D.Q., and M.B. directed the project and edited the manuscript. All authors discussed the results and reviewed the manuscript.

\section*{Acknowledgment}
We acknowledge the support from CMC Microsystems for fabrication services using the 130-nanometre CMOS technology from TSMC. The authors thank Bastien Imbert for assisting in the characterization setup.  

\end{document}